\documentclass[preprint,12pt]{elsarticle}
\usepackage{color}
\pdfoutput=1

\usepackage{graphicx}

\usepackage{amssymb}
\usepackage{amsthm}
\usepackage{hyperref}
\hypersetup{
    pdfnewwindow=true,      
    colorlinks=true,       
    linkcolor=red,          
    citecolor=blue,        
    filecolor=magenta,      
    urlcolor=cyan           
}

\journal{Astroparticle Physics}

\begin{document}

\begin{frontmatter}



\title{Bounding the Time Delay between High-energy Neutrinos and Gravitational-wave Transients from Gamma-ray Bursts}


\author[APC]{Bruny Baret}
\author[Columbia]{Imre Bartos\corref{cor1}}
\author[APC]{Boutayeb Bouhou}
\author[LIGOLab]{Alessandra Corsi}
\author[AEIHannover]{Irene Di Palma}
\author[APC,ParisSud]{Corinne Donzaud}
\author[APC]{V\'eronique Van Elewyck}
\author[Sweden]{Chad Finley}
\author[Cardiff]{Gareth Jones}
\author[APC]{Antoine Kouchner}
\author[Columbia]{Szabolcs M\'arka}
\author[Columbia]{Zsuzsa M\'arka}
\author[APC]{Luciano Moscoso}
\author[APC]{Eric Chassande-Mottin}
\author[AEIHannover]{Maria Alessandra Papa}
\author[IPHC]{Thierry Pradier}
\author[ELTE]{Peter Raffai}
\author[Columbia]{Jameson Rollins}
\author[Cardiff]{Patrick Sutton}

\address[APC]{AstroParticule et Cosmologie (APC), CNRS: UMR7164-IN2P3-Observatoire de Paris-Universit\'e Denis Diderot-Paris VII-CEA: DSM/IRFU, France}
\address[Columbia]{Department of Physics, Columbia University, New York, NY 10027, USA}
\address[LIGOLab]{LIGO Laboratory, California Institute of Technology, Pasadena, CA 91125, USA}
\address[AEIHannover]{Albert-Einstein-Institut, Max-Planck-Institut f\"ur Gravitationsphysik, D-30167 Hannover, Germany}
\address[ParisSud]{Universit\'e  Paris-sud, Orsay, F-91405, France}
\address[Sweden]{Oskar Klein Centre \& Dept. of Physics, Stockholm University, SE-10691 Stockholm, Sweden}
\address[Cardiff]{Cardiff University, Cardiff CF24 3AA, UK}
\address[IPHC]{University of Strasbourg \& Institut Pluridisciplinaire Hubert Curien, Strasbourg, France}
\address[ELTE]{E\"otv\"os University, ELTE 1053 Budapest, Hungary}
\cortext[cor1]{Corresponding author. Email: ibartos@phys.columbia.edu}

\begin{abstract}
We derive a conservative coincidence time window for joint searches of gravita-tional-wave (GW) transients and high-energy neutrinos (HENs, with energies $\gtrsim 100$GeV), emitted by gamma-ray bursts (GRBs). The last are among the most interesting astrophysical sources for coincident detections with current and near-future detectors. We take into account a broad range of emission mechanisms. We take the upper limit of GRB durations as the $95\%$ quantile of the $T_{90}$'s of GRBs observed by BATSE, obtaining a GRB duration upper limit of $\sim 150$s. Using published results on high-energy ($>100$MeV) photon light curves for 8 GRBs detected by Fermi LAT, we verify that most high-energy photons are expected to be observed within the first $\sim 150$s of the GRB. Taking into account the breakout-time of the relativistic jet produced by the central engine, we allow GW and HEN emission to begin up to $100$s before the onset of observable gamma photon production. Using published precursor time differences, we calculate a time upper bound for precursor activity, obtaining that $95\%$ of precursors occur within $\sim 250$s prior to the onset of the GRB. Taking the above different processes into account, we arrive at a time window of  $t_{HEN} - t_{GW}\in[-500\mathrm{s},+500\mbox{s}]$. Considering the above processes, an upper bound can also be determined for the expected time window of GW and/or HEN signals coincident with a detected GRB, $t_{GW} - t_{GRB} \approx t_{HEN} - t_{GRB} \in[-350\mbox{s},+150\mbox{s}]$.
\end{abstract}


\end{frontmatter}



\section{Introduction}

The joint detection of gravitational waves (GWs) and high-energy neutrinos (HENs) from astrophysical sources in a multimessenger approach promises significant scientific benefit over single messenger searches. Such advantages include increased sensitivity, as well as additional information on the underlying mechanisms responsible for creating such multimessenger sources \cite{ET,asoicecube0264-9381-25-11-114039,microquasarHENPradier2009268,2009IJMPD..18.1655V,1742-6596-243-1-012002}.

GW and HEN detectors have made significant progress in recent years. The first generation of GW observatories, such as LIGO \cite{LIGO0034-4885-72-7-076901}, Virgo \cite{VIRGO0264-9381-23-19-S01} and GEO \cite{GEO0264-9381-19-7-321} have already made astrophysically interesting observations (e.g. \cite{PhysRevD.69.082004,PhysRevD.76.062003,SGRGW,1538-4357-683-1-L45,0004-637X-681-2-1419,1538-4357-701-2-L68,2009Natur.460..990A}). The second generation of these detectors (e.g. \cite{advancedLIGO0264-9381-27-8-084006}) is expected to be roughly an order of magnitude more sensitive around $\sim 150$Hz. HEN observatories currently under operation are IceCube \cite{icecube}, a cubic kilometer detector at the geographic South Pole, and ANTARES \cite{ANTARES} in the Mejditerranean sea. IceCube has reached its design volume with 86 strings, while ANTARES is operating at its design volume with 12 strings. ANTARES is planned to be upgraded into a cubic kilometer detector called KM3NET in the following years \cite{KM3NeTdeJong2010}. The Lake Baikal neutrino detector is also planned to be upgraded to a cubic kilometer volume aiming to detect HENs \cite{Aynutdinov200914}.

One of the most promising class of sources for joint GW+HEN searches is represented by gamma-ray bursts (GRBs). GRBs are thought to be produced by internal shocks in relativistic expanding fireballs from energetic cosmic explosions \cite{waxmanbachall}. Fireballs \cite{Fireball}, i.e. the relativistic outflows or jets of plasma, are created when the central engine releases a large amount of energy over a short time and small volume \cite{Fireballshock}. These jets are often thought to be launched along the rotational axis of the progenitor, powered by the gravitational energy released during temporary mass accretion onto the central black hole \cite{HENfromSuccessfulnChokedGRB}.

There are two main cosmic progenitor candidates that are expected to produce the relativistic outflows and shocks of magnetized plasma described above. Long-duration GRBs ($\geq$2s) are thought to be produced by stellar core-collapse \cite{GWsignaturefromCoreCollapse}, while short GRBs ($\leq$2s) are predominantly associated with compact stellar mergers \cite{Nakar06}.

Joint GW+HEN searches are of special importance for sources dark in other messengers, e.g. that have little or no electromagnetic emission. Such potential sources include choked GRBs \cite{chokedfromreverseshockPhysRevD.77.063007}. If a stellar core-collapse emits a relativistic jet from its core that emits gamma-rays, these gamma-rays will only be observable from the outside once the relativistic jet broke out of the stellar envelope. Relativistic jets that stall before reaching the surface of the star can still emit observable HENs \cite{chokedfromreverseshockPhysRevD.77.063007,imre1}, while the envelope is transparent to GWs. The result is GW+HEN emission with little or no gamma-ray counterpart.

Previous single or multimessenger searches for GWs, HENs and/or electromagnetically detected GRBs have defined various different time windows. Aso \emph{et al.} \cite{asoicecube0264-9381-25-11-114039} designed a GW+HEN multimessenger search algorithm and investigated possible time windows from 0.1s to up to 1 day. In a search for HEN counterparts of detected GRBs, Abbasi \emph{et al.} (IceCube Collaboration) \cite{2010ApJ...710..346A} consider three different time windows around each burst. The first time window covers the observed prompt gamma-ray emission of the GRB, expecting prompt neutrino emission to overlap with prompt gamma-ray emission. The second considered time window covers neutrino-producing processes before gamma-rays can escape from the fireball. This time window is taken to be 100s immediately preceding the prompt emission time window. Abbasi \emph{et al.} (IceCube Collaboration) \cite{2010ApJ...710..346A} also consider a generic time window that is chosen to be much wider than the previous two to include possible unknown mechanisms. This third time window is chosen to be $[-1\mbox{h},3\mbox{h}]$ around the start of the prompt emission, a compromise between including unknown mechanisms and keeping the HEN background low.

In a HEN search for GRB 080319B, one of the brightest GRBs ever observed, the IceCube Collaboration \cite{0004-637X-701-2-1721} uses two time windows for HEN detection. A shorter time window of 66s was used that overlapped with the GRB's prompt gamma-ray emission. Another, $\sim300$s long extended time window was also analyzed. Data from the detector was only available in this 300s around the GRB which motivated the choice of this time window.

Very recently a search for neutrinos from 117 GRBs was conducted using data from the IceCube detector's 40 string configuration \cite{2011arXiv1101.1448I}. Two different searches were performed; one model independent search, using no model for the energy distribution of HENs from GRBs. This search used a time window of $\pm24$ hours around the burst. The second, model dependent search used the predicted energy distribution from GRBs by Guetta \emph{et al.} \cite{NeutrinoBATSEGuetta2004429}. This search considered the observed start ($T_{start}$) and stop ($T_{stop}$) times of gamma-ray emission from each GRB. A probability distribution was assigned for expected neutrino times with uniform probability between $T_{start}$ and $T_{stop}$, and with Gaussian tails around this uniform window. The width of the tails was chosen to be $T_{stop} - T_{start}$, and was constrained to at minimum 2 seconds and at maximum 30 \cite{2011arXiv1101.1448I}.

In GW burst searches triggered by long and/or short GRBs, performed with the LIGO-Virgo GW network,  \cite{PhysRevD.72.042002,PhysRevD.77.062004,0004-637X-681-2-1419,0004-637X-715-2-1438}, a time window of $[-120\mbox{s},+60\mbox{s}]$ was used around the GRB trigger. This time window is longer than the duration of most analyzed GRBs. It was also chosen to include models predicting gamma-ray emission up to 100s after initial GW emission (e.g. \cite{fireball0034-4885-69-8-R01}), as well as measurement uncertainties in GRB trigger time. For a search for GW inspiral counterparts of detected short GRBs, Abadie \emph{et al.} (LIGO Scientific Collaboration and Virgo collaboration) \cite{0004-637X-715-2-1453} used a time window of $[-5\mbox{s},+1\mbox{s})$ around the trigger time of GRBs, aiming to capture ``the physical model with some tolerance for its uncertainties.''

The present article, for the first time, discusses the temporal structure\footnote{I.e. the time intervals with possible emission and their relative time of occurrence.} of GW and HEN emission from GRBs relative to each other, with the main goal of defining a common on-source time window. As many aspects of GRB emission processes are still debated, we base our analysis largely on model-motivated comparisons with observations.


The article is organized as follows. We first (Section \ref{discussion}) present a summary of the obtained GW+HEN coincidence time window building on the results and discussion of Section \ref{sec:GRB}. In Section \ref{sec:GRB} we review the potential emission processes in detail, using model-motivated comparisons with observations and simulations to constrain the duration of each emission epoch. We refer the reader to Section \ref{discussion} for an overview and summary of the results, while the details can be found in Section \ref{sec:GRB}.

\section{Summary - Time Window of GW+HEN Emission}
\label{discussion}

In this section we determine an upper bound for the time difference of HEN and GW signals from GRBs. We will refer to this time interval as \emph{GW+HEN coincidence time window}. We aim to obtain an upper bound in which HENs and GW transients are most likely to arrive. The section summarizes the expected durations of different emission processes that are discussed in details in the rest of the article.

The analysis mainly relies on model-motivated comparisons with observations and simulations. We consider different types of information we have on GRBs from their detected electromagnetic radiation, and infer to the plausible time frame of GW or HEN emission based on this information. We use observations by BATSE \cite{BATSE4B}, Swift \cite{Swift} as well as Fermi LAT \cite{0004-637X-697-2-1071}. We estimate the duration of different emission epochs, and combine them to obtain an overall time window. While a joint search for GW and HEN signals mainly aims to detect astrophysical sources with unobserved electromagnetic (EM) counterpart, we constrain their emission based on the emission of detected sources, assuming that these two types have similar emission.

\begin{figure}
\begin{center}
\resizebox{1\textwidth}{!}{\includegraphics{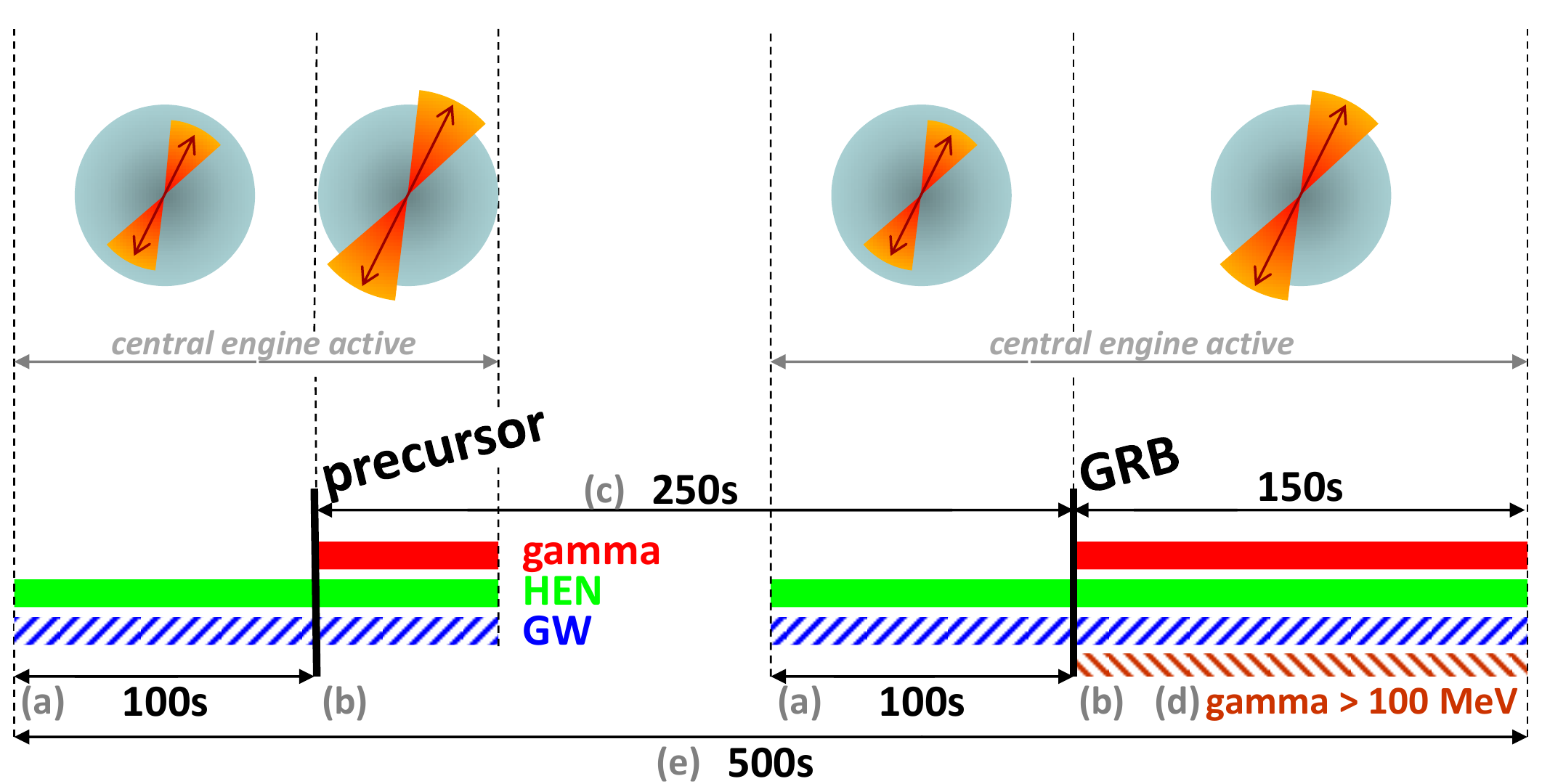}}
\end{center}
\caption{Summary of upper bound of GRB emission process durations taken into account in the total GW+HEN coincidence time window. (a) active central engine before the relativistic jet has broken out of the stellar envelope; (b) active central engine with the relativistic jet broken out of the envelope; (c) delay between the onset of the precursor and the main burst; (d) duration corresponding to $90\%$ of GeV photon emission; (e) time span of central engine activity. Overall, the considered processes allow for a maximum of 500s between the observation of a HEN and a GW transient, setting the coincidence time window to $[-500\mbox{s},500\mbox{s}]$. The time window for GW or HEN signals from the onset of the GRB is $t_{GW} - t_{GRB} \approx t_{HEN} - t_{GRB} \in[-350\mbox{s},+150\mbox{s}]$. Note that we show a period between the end of the precursor emission and the start of the main GRB with no GW or HEN emission. While we cannot exclude the possibility of GW or HEN emission in this period, such emission would have no effect on our estimated time window. The top of the figure shows a schematic drawing of a plausible emission scenario.} \label{fig:emission}
\end{figure}

A summary of the considered emission processes and their time frames are shown in Figure \ref{fig:emission}. Note, that for the case of GW emission, a time frame possibly having GW signals means that there can be one or more short transients at any time of the time frame. There can also be a longer GW emission during this time (see Section \ref{sec:GW} for further details).

\begin{enumerate}
\item[$\bullet$] We start with the onset of gamma-ray emission from the main GRB. The duration of the main GRB is, in $95\%$ of cases, shorter or equal than 150s (see Section \ref{sec:gammaemission}). During this time frame, as the central engine needs to be active, both HEN and GW emission is possible.
\item[$\bullet$] Recent Fermi/LAT observations (e.g. \cite{MeszarosHEG,2009arXiv0907.0715O}) have shown that at least some GRBs also emit high-energy (GeV) gamma-rays, with temporal structure different from that of gamma-rays of typical energies (see Section \ref{sec:highenergygammaemission}). Our estimated time scale of the emission shows that most ($\gtrsim 90\%$) of the high-energy photons are expected to be detected within $150$s after the onset of the GRB.
\item[$\bullet$] GW and HEN emission can commence up to 100s before the onset of gamma-ray emission (see Section \ref{sec:neutrino}) due to the time it takes for the relativistic jet to break out of the star (e.g, \cite{chokedfromreverseshockPhysRevD.77.063007}). The central engine is active during this time, for the jet needs to be actively driven in order to advance.
\item[$\bullet$] Precursor emission can begin up to $t_{95}^{precursor} \simeq 250$s prior to the onset of the main GRB (see Section \ref{sec:precursor}). Since a precursor can be active practically any time within this time frame, and since the activity of the central engine is a prerequisite of precursor emission, both HEN and GW emission is possible during this period.
\item[$\bullet$] The activity of the central engine responsible for a precursor can begin up to 100s before the onset of gamma/X-ray emission by the precursor (see Section \ref{sec:precursor}). During this time, as the central engine is active, both HEN and GW emission is possible.
\item[$\bullet$] We note that our choice for the GW/HEN coincidence window is not affected by what happens in between the precursor(s) and the main GRB. Since HEN and/or GWs may be present both at the precursor and during the main GRB, the GW/HEN coincidence window needs to extend anyway from the precursor up to the end of the main event. Thus, such a choice also covers the case in which GWs and/or HEN are also emitted between the precursor and the main GRB.
\end{enumerate}
Considering the processes described above, we obtain a time frame of $500$s during which both HEN and GW emissions are possible. This means that the difference between the arrival time $t_{HEN}$ of a HEN and the arrival time $t_{GW}$ of a GW transient from the same GRB is within the upper bound (i.e. time window) of
\begin{equation}
t_{HEN} - t_{GW}\in[-500\mbox{s},+500\mbox{s}]
\end{equation}
Note that multiple HEN (or GW) signals from a given source should arrive within a $500$s time window.

We emphasize that the above time window is an estimated upper bound for GRBs. While the emission mechanisms considered are connected to the progenitor associated with long GRBs, short GRBs likely have shorter emission span and are therefore also bound by this time window. Also, we exclude potential GW/HEN emission from GRB afterglows from the time window (the emission of ultra-high energy ($\approx 10^8-10^9$GeV) neutrinos may be more extended; see Section \ref{sec:neutrino}). Other effects, such as some predicted by different Quantum Gravity models (e.g. \cite{2007NatPh...3...87J}), can introduce additional time delay between different messengers. Searching for such effects may require larger time windows \cite{2007NatPh...3...87J}.

The processes considered above can also be used to set an upper bound on time window of GW or HEN signals coincident with an observed GRB. If the onset of a GRB is detected at $t_{GRB}$, both GW and HEN signals are expected to arrive within the time window
\begin{equation}
t_{GW} - t_{GRB} = t_{HEN} - t_{GRB} \in[-350s,+150s]
\end{equation}

\section{Emission Processes in Gamma-ray Bursts}
\label{sec:GRB}

The widely accepted model of long GRBs, the so-called collapsar model, associates long GRBs with the collapse of the core of a massive star, that drives an outgoing relativistic outflow (see, e.g, \cite{chokedfromreverseshockPhysRevD.77.063007}). According to the fireball model of GRBs \cite{Fireball}, gamma-rays, as well as HENs are produced by shocks in this relativistic outflow.

The plasma outflow from the central engine reaches relativistic velocities only after leaving the Helium core of the star \cite{chokedfromreverseshockPhysRevD.77.063007}. Consequently, the time it takes for the outflow to break out of the star is approximately equal to the duration of crossing the Helium core ($t_{He}\approx10-100$s) \cite{chokedfromreverseshockPhysRevD.77.063007} (see also \cite{1538-4357-588-1-L25}). The outflow can advance only as long as it is driven by the central engine, therefore the duration of the outflow has to be longer than $t_{He}$ in order for the outflow to break out of the star. Due to this requirement, there can be stellar core-collapses where the outflow is unable to break through the stellar envelope \cite{ChokedGRBPhysRevLett.87.171102}. In these so-called \emph{choked GRBs}, whose number may exceed the number of successful GRBs \cite{ChokedGRBPhysRevLett.87.171102}, GWs, low-energy neutrinos and potentially HENs can escape the star, while the star remains dark in gamma-rays.

Besides long GRBs, another distinct subclass of gamma-ray bursts are short-duration, hard-spectrum GRBs, with duration $\lesssim 2$s (e.g, \cite{Nakar06}). At least some of these short GRBs are associated with neutrons star - neuron star or neutron star - black hole binary coalescenses (e.g, \cite{Nakar06}). Such binary coalescence constitutes the central engine that create relativistic outflow as discussed above for long GRBs, responsible for the emission of gamma-rays and high energy neutrinos.

This section discusses the different emission processes with special emphasis on their relative timing, defining their contribution to the GW+HEN coincidence time window.

\subsection{Gamma-ray Emission}
\label{sec:gammaemission}

\begin{figure}
\begin{center}
\resizebox{1\textwidth}{!}{\includegraphics{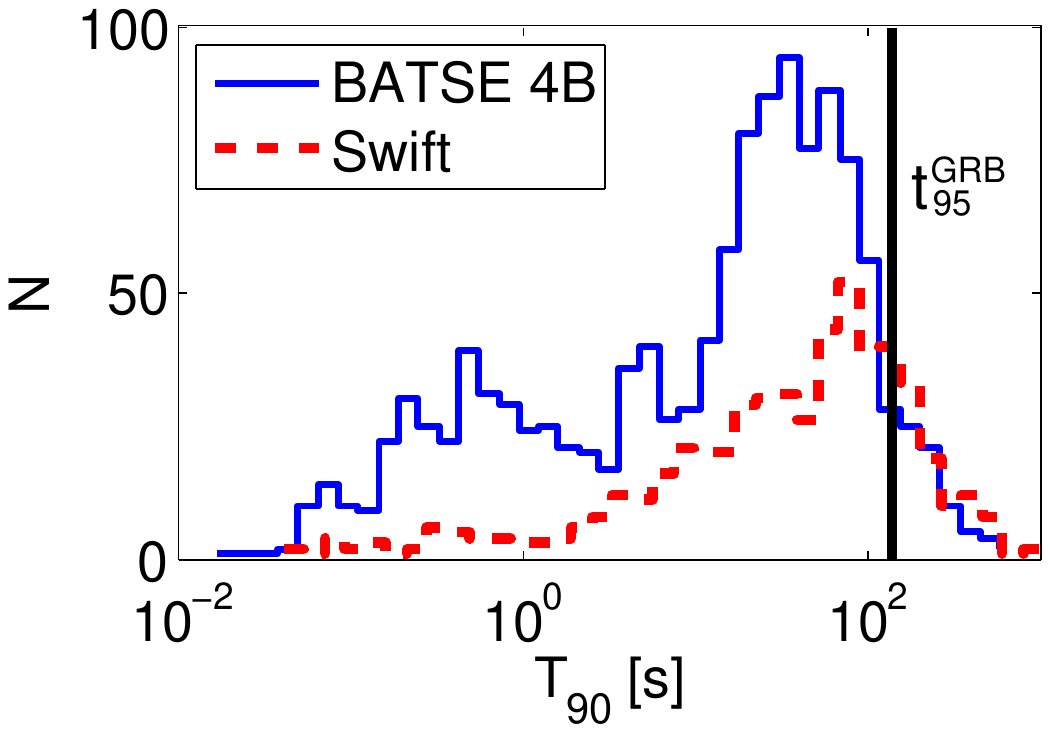}\includegraphics{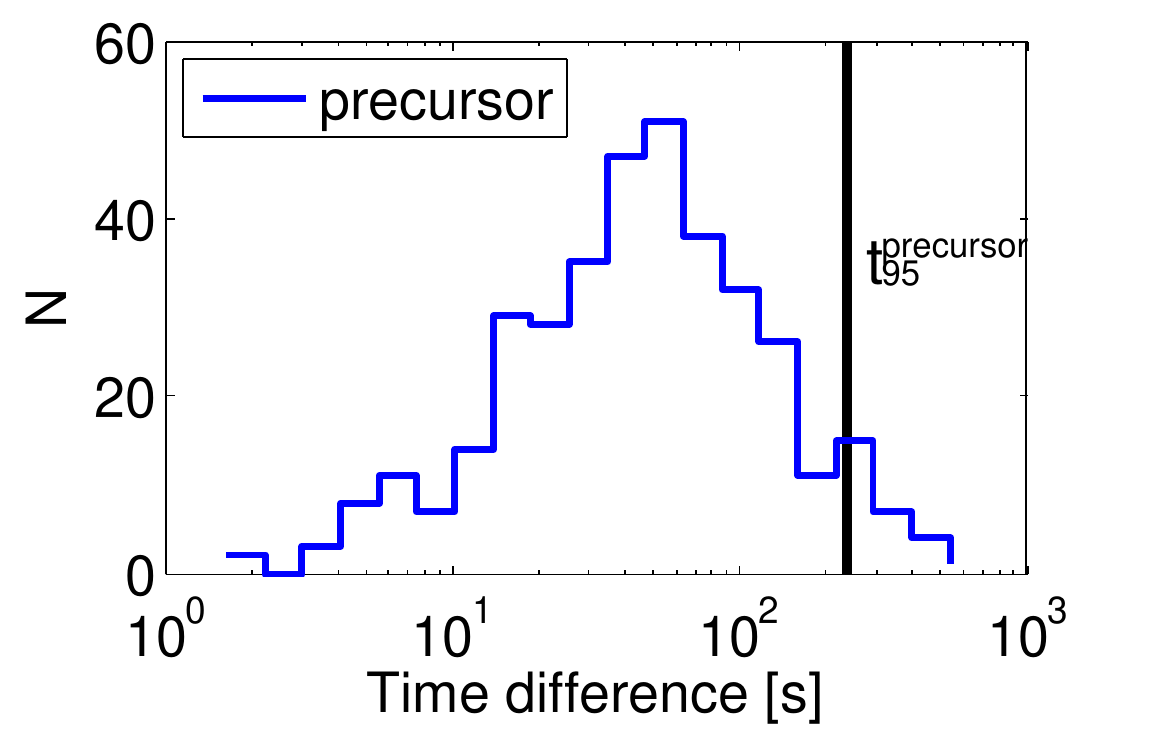}}
\end{center}
\caption{(left) Distribution of T$_{90}$'s for GRBs measured by BATSE (straight line, 1234 GRBs, \cite{BATSE4B}) and Swift (dotted line, 534 GRBs, \cite{Swift}). The vertical line shows $t_{95}^{GRB}$, i.e. the 95$\%$ quantile of the $T_{90}$'s of BATSE GRBs. (right) Distribution of time difference between the onset of a precursor and the onset of the main GRB. The vertical line shows $t_{95}^{precursor}$, i.e. the 95$\%$ quantile of the time differences (the delays were obtained by Burlon \emph{et al.} \cite{PrecursorBATSE} using data from BATSE \cite{BATSE4B}).} \label{fig:t95}
\end{figure}

The observed temporal structure of gamma-ray emission from GRBs can be used as the foundation in the description of the temporal structure of GW and HEN emission. For the purposes of GW+HEN analysis, we define a practical upper limit for GRB duration as the 95$\%$ quantile of the $T_{90}$'s of GRBs detected by the Burst Alert and Transient Source Experiment (BATSE). We obtain this limit using data from the 4th BATSE GRB catalog (1234 GRBs, \cite{BATSE4B}), that includes 1234 GRBs with duration information (see Figure \ref{fig:t95}). As the measure of GRB durations, we use their T$_{90}$, the time interval in which the integrated photon counts from a GRB in the BATSE detector increased from 5$\%$ to 95$\%$. See Figure \ref{fig:t95} for the distribution of T$_{90}$'s of BATSE GRBs in comparison with $t_{95}^{GRB}$. The obtained duration upper limit is
\begin{equation}
t_{95}^{GRB} \simeq 150 \enspace \mbox{s}
\end{equation}
For comparison we note here that the $90\%$ and $99\%$ quantiles are $t_{90}^{GRB} \simeq 100$s and $t_{99}^{GRB} \simeq 300$s. Considering a similar analysis for GRBs detected by the Swift satellite (534 GRBs, \cite{Swift}), we find that $87\%$ of Swift GRBs have $T_{90}\lesssim 150$s. For Swift, the $90\%$, $95\%$ and $99\%$ quantiles are 180s, 300s and 500s, respectively. See Figure \ref{fig:t95} for the distribution of Swift T$_{90}$'s.

\subsection{High-energy Gamma-ray Emission}
\label{sec:highenergygammaemission}

While typical gamma-ray energies from GRBs are in the range of $0.1-1$~MeV, some GRBs also emit gamma-rays with energies above $100$~MeV (e.g. \cite{MeszarosHEG,2009arXiv0907.0715O}). Such high-energy component was first observed by the Energetic Gamma-Ray Experiment Telescope (EGRET, e.g, \cite{springerlink:10.1007/BF00658613,1995ARA&A..33..415F,1994ApJ...422L..63S}), and has recently been identified for more than 20 GRBs \cite{2010ApJS..188..405A} by the Large Area Telescope (LAT) on the Fermi satellite \cite{0004-637X-697-2-1071}. GeV gamma-ray emission from these GRBs, in most cases, shows radically different characteristics \cite{MeszarosHEG,2009arXiv0907.0715O,2010MNRAS.403..926G} compared to MeV gamma-ray emission.

The different temporal and spectral behavior of GeV photons imply that they are likely produced by different processes during a GRB than MeV photons. There are different proposed mechanisms responsible for the creation of GeV photons. It is also possible that different mechanisms are responsible for the GeV photon output for different GRBs. Studying the high energy emission of GRB 090510, Ghirlanda \emph{et al.} \cite{Ghirlanda} suggested that this may be explained entirely due to external shock emission. On the other hand, several other authors \cite{2041-8205-709-2-L146,0004-637X-720-2-1008,2010arXiv1009.1432H} have shown that internal shocks may also play a role, especially during the first few seconds of the high-energy tail. Toma \emph{et al.} \cite{2010arXiv1002.2634T} argue that GeV photon emission can be due to Compton upscattered photospheric emission, during which photons are upscattered by electrons in internal shocks. Photohadronic interactions in proton-dominated bursts have also been considered, that may be responsible for the production of the high-energy component \cite{0004-637X-721-2-1891,0004-637X-699-2-953}. In this case, GRBs with GeV photon emission can be excellent sources of HENs. In the more standard Waxman-Bahcall model, however, GeV emission, associated with higher Lorentz factor, results in decreased HEN emission (e.g. \cite{0004-637X-701-2-1721}). See e.g. \cite{MeszarosHEG} for a summary of possible emission mechanisms.

Due to the small number of detected LAT GRBs, as well as our limited understanding of the emission mechanisms, investigating the temporal behavior of high-energy gamma-ray emission is difficult. It seems, however, that in many cases the light curves follow a power-law decay in time: $F_{LAT}\propto t^{-\alpha}$ with $\alpha\simeq1.5$ \cite{2010MNRAS.403..926G,03272009,1538-4357-706-1-L138}, while the rising parts for some cases are consistent with $F_{LAT}\propto t^2$. We note that the characteristic rising and decay were observed both for short and long bursts \cite{2010MNRAS.403..926G}.

For the purposes of GW+HEN analysis, we are interested in the proportion of high-energy photons emitted in a given time window, $T$, following the onset of the GRB. For this estimate, following \cite{Ghirlanda}, we model the temporal distribution of high-energy photons with a smoothly broken power law
\begin{equation}
F_{LAT}(t) = \frac{A(t/t_b)^\alpha}{1+(t/t_b)^{\alpha+\beta}}+B
\label{flat}
\end{equation}
with $\alpha=2$ and $\beta=1.5$ as discussed above. The peak emission is at time $t_{peak} = t_b(\alpha/\beta)^{1/(\alpha+\beta)}$. $A$ is an amplitude and $B$ is the background level, both irrelevant for the calculation of the temporal behavior (for simplicity, we will assume $B=0$).

Due to the low number of available data, we restrict our analysis to estimating the ratio of GeV photons arriving in the 150s time window, i.e. the GRB emission time window described above in Section \ref{sec:gammaemission}. Following \cite{2010MNRAS.403..926G}, we use 8 out of the 12 GRBs for which the detected GeV emission has high enough signal-to-noise ratio for the reconstruction of the temporal structure. We assume that the emission of these GRBs follow Equation \ref{flat}, with $t_{peak}$ equal to the maximum shown in Figure 4. in \cite{2010MNRAS.403..926G}. Taking all 8 GRBs into account with equal weights, we verify that $\sim90\%$ of the GeV photons arrive within a 150s time window after the onset of the GRB (see Figure \ref{fig:emission}). For comparison, $50\%$ of the photons arrive within $\sim7$s, while $95\%$ of the photons arrive within $650$s. We conclude that the 150s time window is a reasonable upper bound for the high-energy gamma-ray emission as well. We note that the obtained time window does not depend significantly on the initial rising part. The important factors are the peak time and the slope afterward.

We note that, while the majority of GeV photons seems to arrive within the first 150s after the onset of the GRB, the importance of detected GeV photons from the perspective of neutrino emission can be that it might indicate a process, such as external shocks, that can produce HENs. If in such process the emission of GeV photons is not proportional to the emission of HENs (i.e. if the durations of the arrival of $\sim90\%$ of GeV photons and HENs are different), the duration of HEN emission might be longer than 150s.



\subsection{Gravitational-wave Emission}
\label{sec:GW}

While gamma-ray and HEN emission from GRBs are practically independent from the specific mechanisms driving the (varying) relativistic outflow\footnote{The distance from the central engine at which observable gamma-rays and HENs are produced is much larger than the central engine (e.g. \cite{waxmanbachall}).}, GW emission is closely connected to the central engine.

Various mechanisms can result in the emission of GW transients. Here we first consider stellar core collapses, the likely progenitors of at least some long GRBs. For these the collapse of the rotating iron core, as well as the following rebound in the inner core instantly produce a GW burst lasting for tens of milliseconds (see e.g. \cite{GWsignaturefromCoreCollapse}). Other mechanisms following core bounce can also emit GWs shortly, within a few seconds after the collapse. These processes include rotational non-axisymmetric instabilities, post-bounce convection, standing-accretion-shock instability (SASI), or non-radial proto-neutron star pulsation \cite{GWsignaturefromCoreCollapse}.

The collapse results in the formation of a protoneutron star, that can further collapse to form a stellar-mass black hole (e.g. \cite{2010arXiv1010.5550O} and references therein). The rotating, collapsing core might also fragment into two or more compact objects, whose coalescence can emit strong GWs \cite{0004-637X-565-1-430}. After the formation of a black hole, weaker GW emission is possible from accreting matter around this central black hole that can fragment due to gravitational instability. The resulting black hole - fragment binaries can emit GWs detectable by Earth-based GW interferometers \cite{1538-4357-579-2-L63,1538-4357-630-2-L113,0004-637X-658-2-1173}. For high enough accretion disk masses, the self-gravity of the disk can result in gravitational instabilities such as spiral arm or bar formation, emitting GWs (e.g. \cite{Putten03082002}, \cite{0004-637X-589-2-861} and references therein). Non-uniform, non-axisymmetric accretion can also deform the central black hole that will emit gravitational waves during its ring-down \cite{0004-637X-589-2-861}. Such GW emission processes can last for the duration of the accretion, i.e. possibly the duration of the GRB.

Compact binary coalescence, the likely progenitor of short GRBs, is anticipated to be strong GW emitters in the sensitive frequency band of Earth-based GW detectors (e.g \cite{0004-637X-715-2-1453,2009A&A...498..329G,0004-637X-589-2-861}). Double neutron stars, black hole-neutron star binaries and black hole-white dwarf binaries are considered as possible sources (e.g. \cite{0004-637X-589-2-861}). Most of the GW output of such sources is emitted in the form of a very short, $\sim1$ms long transient (e.g. \cite{PhysRevD.79.024003,0004-637X-589-2-861}). Furthermore, many short GRBs occur at low redshifts \cite{1538-4357-643-2-L91}, increasing their significance as potential GW (and HEN) sources.

Based on the models discussed above, GRBs are expected to be strong emitters of GWs. While core-collapse and binary merger are expected to emit the strongest GW transient, emission is possible during the entire duration when the central engine is active (see Figure \ref{fig:emission}). GW emission can consist of either one or more short burst (e.g. due to core collapse or the infalls of fragments of the accretion disk \cite{1538-4357-579-2-L63,1538-4357-630-2-L113,0004-637X-658-2-1173}), or longer duration GWs (e.g. due to the emission of rotational energy \cite{Putten03082002}). If the central engine consists of an accretion disk, the activity of the central engine coincides with the emission of GWs.

\subsection{High-energy Neutrino Production}
\label{sec:neutrino}

In a GRB, HENs and gamma-rays are expected to be produced due to the variability of the central engine's activity, that results in fluctuations of the relativistic outflow, creating internal shocks in the ejecta (e.g. \cite{0004-637X-723-1-267,0004-637X-698-2-1261,NeutrinoBATSEGuetta2004429,0004-637X-559-1-101}). These internal shocks accelerate electrons and protons in the outflow through the process of Fermi acceleration. Shock-accelerated electrons radiate their energy through synchrotron or inverse-Compton radiation (e.g. \cite{0004-637X-698-2-1261,astro_sources_neutrinos}), emitting gamma-rays. Shock-accelerated protons interact with gamma-rays ($p\gamma$) as well as with other, non-relativistic protons ($pp$), producing charged pions and kaons \cite{HEN2005MPLA...20.2351R}. Pions and kaons from these processes decay into HENs through \cite{HENAndoPhysRevLett.95.061103}
\begin{equation}
\pi^{\pm}, K^{\pm} \rightarrow \mu^{\pm} + \nu_{\mu}(\overline{\nu}_{\mu}) 
\label{piondecay}
\end{equation}
The resulting muons from these interactions are expected to immediately undergo radiative cooling, making this secondary HEN production insignificant \cite{HENAndoPhysRevLett.95.061103,HEN2005PhysRevLett.93.181101,chokedfromreverseshockPhysRevD.77.063007}. For the proton energy range $E_p\approx 10^4-10^{5.2}$~GeV the $p\gamma$ process dominates HEN production, while outside of this energy range the $pp$ process is dominant \citep{HEN2005MPLA...20.2351R}.

Since internal shocks in the relativistic outflow result in both gamma-ray and HEN emission, HENs are expected to be produced during the emission of gamma-rays (see Figure \ref{fig:emission}).

For the case of long GRBs created by core-collapse supernovae, internal shocks in the relativistic outflow can occur even before the outflow emerged from the stellar envelope, therefore HENs are also expected to be produced before observable gamma-ray emission \cite{chokedfromreverseshockPhysRevD.77.063007,precursorneutrinos}. While gamma rays emitted in this early phase cannot escape from the star due to its large optical depth, neutrinos have much longer mean free paths and may therefore pass through the stellar envelope. This pre-GRB neutrino emission is expected to precede the start of gamma-ray emission by up to 100s \cite{chokedfromreverseshockPhysRevD.77.063007,precursorneutrinos} (see Figure \ref{fig:emission}). 

GRB afterglows are not considered here as a part of the coincidence time window. Afterglows are produced by the relativistic jet driven into the surrounding medium (e.g. \cite{astro_sources_neutrinos}). The observed radiation, similarly to the case for prompt gamma-ray emission, is produced by synchrotron emission of shock-accelerated electrons. The energy distribution of protons is expected to be similar to that of electrons \cite{astro_sources_neutrinos}, therefore the softer emission spectrum of afterglows indicate that protons might also be of lower energy. These protons would need higher energy gamma-rays to produce neutrinos\footnote{In photomeson interactions producing HENs, the photon's energy $\epsilon_\gamma$ and the proton's energy $\epsilon_p$ are related, at the threshold of the $\Delta$-resonance, by
\begin{equation}
\epsilon_\gamma\epsilon_p = 0.2 \mbox{GeV}^2 \Gamma^2
\end{equation}
in the observer frame, where $\Gamma$ is the bulk Lorentz factor of the outflow \cite{astro_sources_neutrinos}.}, that are scarcely present in the afterglow spectrum.

We note that, while HEN production in GRB afterglows is likely not significant, a few ultra high-energy neutrinos (UHENs) of energies $\sim10^8-10^9$GeV might be emitted during the afterglow phase \cite{UHEN,astro_sources_neutrinos,UHEN2}, if GRBs can accelerate some protons to energies $\epsilon_p \sim 10^{11}$GeV. In our time window estimation, we do not take into account such emission from the afterglow.

\subsection{GRB Precursors}
\label{sec:precursor}

Gamma-ray bursts are sometimes preceded by fainter, softer electromagnetic emissions, so-called precursors (e.g. \cite{PrecursorBATSELazzati}). The underlying mechanism(s) creating these precursors is yet unknown. They have been detected for about $8-20\%$ of GRBs \cite{PrecursorBATSELazzati,precursorSWIFTBurlon,PrecursorBATSE,2010arXiv1009.1385T}, with some GRBs apparently having multiple precursors. It is expected, however, that a higher percentage of GRBs are preceded by precursors, as many of these might be missed due to, e.g., beaming, low signal-to-noise ratio, the proximity of the precursor and the main event, or the definition of what is considered a precursor \cite{PrecursorBATSELazzati}.

Analyzing precursors detected by BATSE, Lazzati \cite{PrecursorBATSELazzati} finds that the net count of a precursor, a value connected to the energy of the events, is up to about $1\%$ of the main burst, most precursors having non-thermal spectra. Burlon \emph{et al.}, using a more detailed analysis of data from the SWIFT telescope \cite{precursorSWIFTBurlon} and BATSE \cite{PrecursorBATSE}, finds that precursors, on average, emit $\sim30\%$ (for SWIFT) and $\sim10-20\%$ (for BATSE) of the energy of the main burst. Burlon \emph{et al.} find that precursors and main events have very similar spectral properties, concluding that precursors and main GRBs are likely to be produced by the same underlying mechanism.

There are several different GRB precursor models in the literature (e.g. \cite{PrecursorBATSE}). The ``two-step engine'' model \cite{PrecursorMeszaros} argues that precursors are produced by initial weak jets, that are created by and coincide with the actual core collapse. The ``progenitor precursor'' model (e.g., \cite{PrecursorLazzati}) connects precursors with isotropic emission marking the jet breakout. The ``fireball precursor'' model \cite{PrecursorLi} considers a simple model of an isotropic, post-acceleration GRB with constant rest mass and kinetic energy. Depending on the considered model, either or both the precursor and main GRB can have GW and/or HEN emission. We note that, while the models above are concerned with precursors from stellar core collapses, precursor emission has also been identified from short GRBs \cite{2010arXiv1009.1385T}.

We estimate an upper bound for the time difference between the onset of a precursor and the onset of the main burst using the results of Burlon \emph{et al.} \cite{PrecursorBATSE}. Burlon \emph{et al.} analyzed 2121 BATSE GRBs, out of which 264 ($12.5\%$) was found to be preceded by one or more precursors (in total 369 precursors). From this data, we determine an upper bound $t_{95}^{precursor}$ to the time difference between the start of the precursor and the main burst, which is the $95\%$ quantile of precursor time differences measured between the onset of the precursor and of the main GRB, as calculated by Burlon \emph{et al}. Figure \ref{fig:t95} shows the distribution of the time differences and the $95\%$ quantile, for which we obtain (also see Figure \ref{fig:emission})
\begin{equation}
t_{95}^{precursor} \simeq 250 \enspace \mbox{s}
\end{equation}
For comparison, we note here that $t_{90}^{precursor} \simeq 150$s and $t_{99}^{precursor} = t_{max}^{precursor}\simeq 350$s. We use $t_{95}^{precursor}$ as the upper limit of time difference between the main burst and a precursor.

For the main GRB above, we used an upper limit of $\sim100$s for the time difference between the activation of the central engine and the onset of observable gamma-ray emission. We will use the same delay upper limit for precursors, i.e. the delay of the onset of gamma/X-ray emission from the precursor is estimated to be less than 100s after the start of the central engine (see Figure \ref{fig:emission}).
\newline
\newline
{\bf Acknowledgement}
\newline
\newline
The authors are grateful to John Cannizzo, Peter M\'esz\'aros, Christian Ott, Tsvi Piran, Richard O'Shaughnessy and Eli Waxman for the helpful discussions and valuable comments.





\bibliographystyle{model1-num-names}
\bibliography{References-Neutrino_01}







\end{document}